\documentclass[reprint,superscriptaddress,amsmath,amssymb,aps,prb,showpacs]{revtex4-1}
\usepackage{graphicx}

\begin{document}

\title{Beyond double-resonant Raman scattering:\\ UV Raman spectroscopy of graphene, graphite and carbon nanotubes}

\author{Christoph Tyborski}
\email{Christoph.Tyborski@physik.tu-berlin.de}
\affiliation{Institut f\"{u}r Festk\"{o}rperphysik, Technische Universit\"{a}t Berlin, Hardenbergstra\ss e 36, 10623 Berlin}

\author{Felix Herziger}
\affiliation{Institut f\"{u}r Festk\"{o}rperphysik, Technische Universit\"{a}t Berlin, Hardenbergstra\ss e 36, 10623 Berlin}

\author{Roland Gillen}
\affiliation{Institut f\"{u}r Festk\"{o}rperphysik, Technische Universit\"{a}t Berlin, Hardenbergstra\ss e 36, 10623 Berlin}

\author{Janina Maultzsch}
\affiliation{Institut f\"{u}r Festk\"{o}rperphysik, Technische Universit\"{a}t Berlin, Hardenbergstra\ss e 36, 10623 Berlin}

\date{\today}

\begin{abstract}
We present an analysis of deep-UV Raman measurements of graphite, graphene and carbon nanotubes. For excitation energies above the strong optical absorption peak at the $M$ point in the Brillouin zone ($\approx 4.7\,\text{eV}$), we partially suppress double-resonant scattering processes and observe the two-phonon density of states of carbon nanomaterials. The measured peaks are assigned to contributions from LO, TO, and LA phonon branches, supported by calculations of the phonon dispersion. Moreover, we gain access to the infrared-active $E_{1u}$ mode in graphite. By lowering the excitation energy and thus allowing double-resonant scattering processes, we demonstrate the rise of the \textit{2D} mode in graphite with ultra-short phonon wave vectors.
\end{abstract}

\pacs{63.22.Rc, 78.30.-j, 61.46.Fg, 78.67.Ch}

\maketitle

Graphite, graphene and carbon nantoubes (CNT) have experienced increasing interest in fundamental research in the last decade. In this context, Raman spectroscopy has been established as a powerful experimental technique, since it provides access to both the electronic and vibrational properties of carbon materials \cite{Cardona2007, Jorio2005}. Due to its high sensitivity, it is possible to probe properties like the crystallographic orientation of graphene \cite{Mohjuddin2009, Huang2009}, the number of graphene layers \cite{Herziger2012}, doping \cite{Pisana2007} and strain \cite{Mohjuddin2009, Mohr2012, Narula2012}, as well as the diameter and chiral indices $(n,m)$ of CNT \cite{Telg2004}.

In general, the Raman spectra of graphene, graphite or CNT in the two-phonon region are dominated by double-resonant (DR) Raman modes \cite{Thomsen2000, Maultzsch2004, Venezuela2011}. Especially in single-layer graphene, the prominent \textit{2D} mode outperforms the intensity of the first-order \textit{G} mode by a factor of up to five. \cite{Ferrari2006} However, these strong DR Raman modes are very unique only for graphitic materials. In common semiconductors, away from optical resonances, the second-order spectrum typically shows the two-phonon density of states \cite{Cardona1982}. A possible route to observe the two-phonon density of states also in graphene, graphite, and CNT is the suppression of DR Raman modes. Here, Raman spectroscopy with photon energies in the ultraviolet (UV) spectral range seems most promising, as the strong optical absorption around the $M$ point in the Brillouin zone is then suppressed. Thus, all so-called 'inner' DR scattering processes are selectively inhibited.

In this work, we investigate the Raman process in graphene, graphite, and CNT under UV excitation. For excitation energies well above the $M$-point transition energy of approximately 4.7\,eV, we can selectively suppress the dominant Raman processes that are commonly identified with 'inner' DR Raman scattering. In these cases the two-phonon density of states (pDOS) is observed. Therefore, we gain access to phonon frequencies at high-symmetry points in the Brillouin zone, in particular the infrared-active $E_{1u}$ mode in graphite, which are otherwise not accessible in either first-order or DR Raman scattering in the visible optical range. By lowering the excitation energy towards the $M$-point transition energy, we can initiate inner double-resonance processes and therefore demonstrate the onset of the \textit{2D} mode in graphite. Our interpretations are supported by calculations of the pDOS for graphene, graphite, and CNT. Furthermore, we calculate Raman spectra for graphite in the UV range and observe good agreement with our experimental data.  

\begin{figure}[t]
\includegraphics{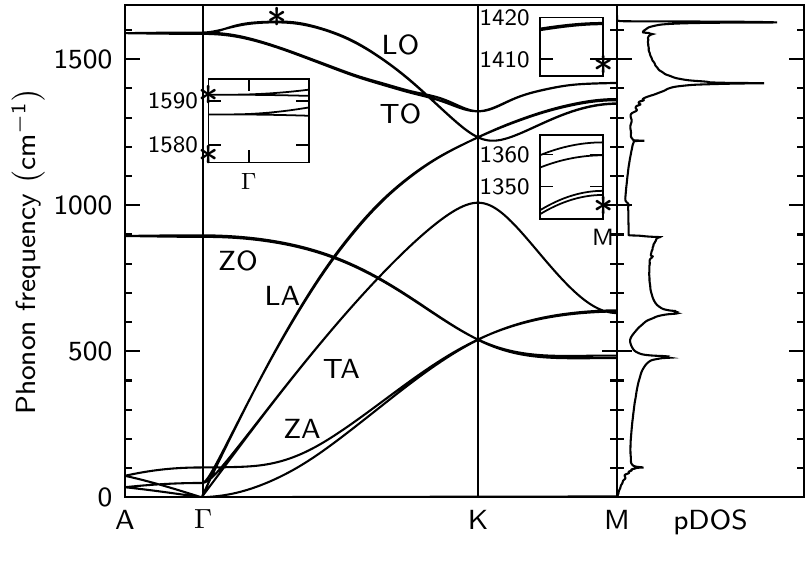}
\caption{Calculated phonon dispersion and the phonon density of states (pDOS) of graphite. Insets show the LO/TO phonon branches at $\Gamma$ ($E_{1u}$ and $E_{2g}$ modes, which both split into LO and TO at finite wave vectors) and the LA, LO/TO phonon branches at $M$, respectively.\label{fig1} Stars mark experimental results.}
\end{figure}

\begin{figure*}[t]
\centering
\includegraphics{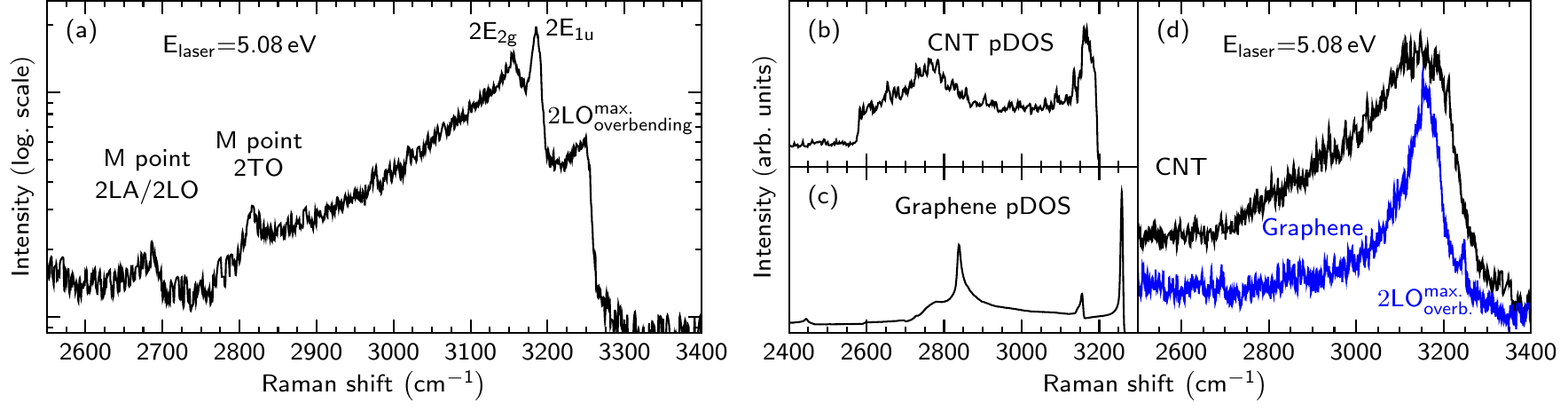}
\caption{(a) Raman spectrum of graphite with an excitation energy of $E_{\text{L}}=5.08\,\text{eV}$ on a logarithmic scale. Peaks are assigned by comparison to maxima in the pDOS (compare Fig.\,\ref{fig1}). (b) and (c) Calculated two-phonon (overtone) pDOS of 40 different CNT and of graphene between $2400\,\text{cm}^{-1}$ and $3300\,\text{cm}^{-1}$. (d) Raman spectra of a HiPCO CNT sample (black) and graphene (blue) with an excitation energy of $5.08\,\text{eV}$.\label{fig2}}
\end{figure*}

We used a Horiba T64000 spectrometer equipped with an Argon ion laser, providing second-harmonic generation of all fundamental laser lines. All spectra are calibrated via the Raman-active vibrational mode of molecular nitrogen ($^{14}\text{N}_{2}$) from the atmosphere. We measured exfoliated graphite, CVD-grown graphene on $\text{SiO}_{2}/\text{Si}$, and a HiPCO-produced buckypaper CNT sample with a diameter range of 7\,\AA\,\,to 13\,\AA. In order to avoid degradation of our samples, we integrated Raman signals of graphene over a large area ($80\times80\,\mu\text{m}^{2}$) with an integration time of $80\,\text{s}$ per spectra and laser powers below $1\,\text{mW}$. Integration times for graphite were in the order of $1.5\,\text{h}$. Measurements were performed using a $40\times$ deep-UV objective with a numerical aperture of $\text{NA}=0.5$, resulting in a laser spot size of $\sim1\,\mu\text{m}$.

Density functional theory calculations of the phonon spectra of graphite and graphene were performed on the level of the local density approximation as implemented in the Quantum ESPRESSO suite \cite{quantumespresso}. The electrons in the system were modeled by projector augmented waves with cutoff energies of 80\,Ry for the electronic wavefunctions. Reciprocal-space integration for the ground state was performed by grids of $21\times21\times1$ $k$ points for graphene and $21\times21\times6$ $k$ points for graphite. We fully optimized the atomic positions and cell parameters of the considered systems until the interatomic forces and the cell pressures were smaller than 0.001\,eV/\AA{} and 0.001\,GPa, respectively. Interactions of the slabs with residual periodic images were minimized by maintaining vacuum layers of at least 25\,\AA{}. Phonon dispersion relations were calculated through density functional perturbation theory using the same $k$-point samplings as before. The phonon density of states was then obtained by Fourier interpolation of the calculated spectra onto a denser grid of $500\times500\times1$ points ($500\times500\times10$ for graphite) and applying a Gaussian broadening of $1\,\text{cm}^{-1}$. 

Figure~\ref{fig1} shows the calculated phonon dispersion of graphite with its density of states (pDOS). The insets illustrate regions with a high pDOS originating from the LO (longitudinal optical), TO (transverse optical), and the LA (longitudinal acoustic) phonon branches at the high symmetry points $\Gamma$ and $M$, respectively. The Raman-active $\Gamma$-point vibration at $\sim1586\,\text{cm}^{-1}$ ($E_{2g}$ irreducible representation) is associated with the \textit{G} band in graphene. In the case of graphite, the single-layer $E_{2g}$ mode gives rise to two modes with the irreducible representations $E_{2g}$ and $E_{1u}$, of which the latter is infrared active. Its frequency is $\omega_{E_{1u}}\sim1591\,\text{cm}^{-1}$ and therefore slightly higher than the Raman-active mode $\omega_{E_{2g}}$ at $\sim1586\,\text{cm}^{-1}$.

Figure~\ref{fig2}\,(a) shows a Raman spectrum of graphite at an excitation energy of $E_{\text{L}}=5.08\,\text{eV}$. In contrast to previous works on UV Raman spectroscopy on graphene and graphite \cite{UV2009,UV2015}, we observe a distinct Raman signal in the two-phonon region from our samples. In fact, we observe five peaks that we assign to the two-phonon density of states. In detail, we observe overtone bands from the $M$ point (LA/LO-derived and TO-derived phonon branches), the overtones of the $E_{2g}$ and $E_{1u}$ modes at the $\Gamma$ point and the overtone of the maximum from the LO-branch overbending near the $\Gamma$ point. The infrared active $E_{1u}$ mode in the second-order spectrum can be observed, since the decomposition of the direct product $E_{1u}\,\otimes\,E_{1u}$ always contains the fully symmetric representation. Assignments are done via comparison of the Raman spectrum with the calculated pDOS, as indicated in Fig.~\ref{fig2}\,(a). The intensity distribution in the pDOS does not fully reflect the intensity distribution in the measured Raman spectrum since coupling matrix elements are not considered. For instance, the electron-phonon coupling of the LO phonon at the $\Gamma$ point is large compared to LO-derived phonons with wavevectors $q\neq\,0$.\cite{Venezuela2011} Thus, the calculated high intensity of the LO-overbending pDOS is not observed experimentally. Instead, in the experiment, the $\Gamma$-point contribution is dominant in Fig.~\ref{fig2}\,(a). We find the maximum of the LO overbending at $1626\,\text{cm}^{-1}$. The frequencies of the $E_{1u}$ and $E_{2g}$ $\Gamma$-point vibrations deduced from our experiment are $1592\,\text{cm}^{-1}$ and $1578\,\text{cm}^{-1}$, respectively, in good agreement with previous theoretical and experimental results (Fig.~\ref{fig1}) \cite{Nemanich1977, Giura2012}. From the measured spectrum in Fig.~\ref{fig2}\,(a), we further find the $M$-point frequency of the TO-derived phonon branch to be $1408\,\text{cm}^{-1}$, in good agreement with calculations. The peak at $2688\,\text{cm}^{-1}$ cannot be distinguished between the LO- and LA-derived branches, since they are very close in frequency at the $M$ point. However, we can determine the upper bound of the LA-phonon branch in graphite to be $1344\,\text{cm}^{-1}$. Thus, the phonon frequencies from the $M$ point and the $E_{1u}$ vibration at the $\Gamma$ point can be investigated by optical spectroscopy, enabling fine adjustments for theoretical calculations of graphene and related systems. 

In Figure~\ref{fig2}\,(b)-(d) we show a comparison of Raman spectra ($E_{\text{L}}=5.08\,\text{eV}$) from graphene and CNT with their calculated two-phonon pDOS.  In contrast to graphite, the experimental Raman spectrum of graphene has a reduced complexity, \textit{i.e.}, less peaks are observed. The peaks at $3162\,\text{cm}^{-1}$ and $3246\,\text{cm}^{-1}$ are attributed to the second-order $E_{2g}$ $\Gamma$-point vibration and the second-order LO-branch overbending, respectively, in analogy to graphite. This is in good agreement with DFT calculations, where we find the peaks at $3154\,\text{cm}^{-1}$ and $3258\,\text{cm}^{-1}$. Compared to graphene and graphite, the UV Raman spectrum of CNTs only consists of a single broad, asymmetric peak around $3150\,\text{cm}^{-1}$. This experimental result is also reflected in the calculated pDOS for the CNT ensemble in Fig~\ref{fig2}\,(b) and can be understood from the fact that phonon frequencies in CNTs sensitively depend on the nanotube diameter and chiral angle \cite{Dobar2003}. The broad diameter distribution of our CNT ensemble (7\AA\,\,to 13\AA) directly results in a broad range of phonon frequencies. For instance, we infer a range of $2\Delta\omega = 80\,\text{cm}^{-1}$ for the second-order high-energy mode frequencies \cite{polsym}, well explaining the experimentally and theoretically observed broad peak centered at $3150\,\text{cm}^{-1}$. In contrast to graphene or graphite, we do not observe distinct peaks in our experimental Raman spectrum that are related to the LO overbending or the pDOS at the M point. Again, also these phonon frequencies show a dependence on tube diameter and chiral angle \cite{Dobar2003}. The expected range of LO-phonon maxima in our CNT sample is $\Delta\omega_{\text{LO,max}}=10\,\text{cm}^{-1}$.\cite{polsym} Thus, all contributions will add up to a broad shoulder on the high-frequency side of the main peak. The broad Raman signal towards lower wavenumbers is attributed to LO-, LA-, and TO-derived phonon bands from the M point.

\begin{figure}
\includegraphics{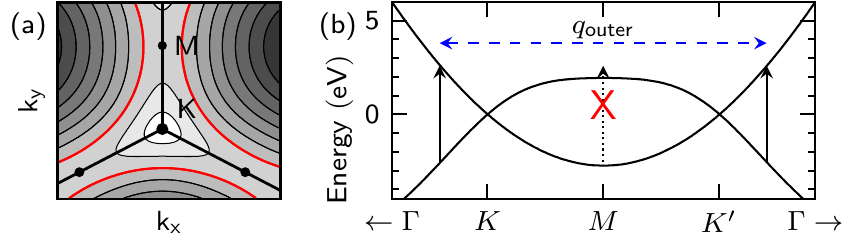}
\caption{(a) Equi-energy contours ($E_{c}-E_{v}=\text{const.}$) from the electronic band structure of graphene. Highlighted is the contour with $E_{c}-E_{v}=5.08\,\text{eV}$. (b) Schematic view of a DR scattering process with an excitation energy of $5.08\,\text{eV}$. Only 'outer' processes can be double resonant above the transition energy at the $M$ point. \label{fig3}}
\end{figure}

\begin{figure}
\includegraphics[width=\columnwidth]{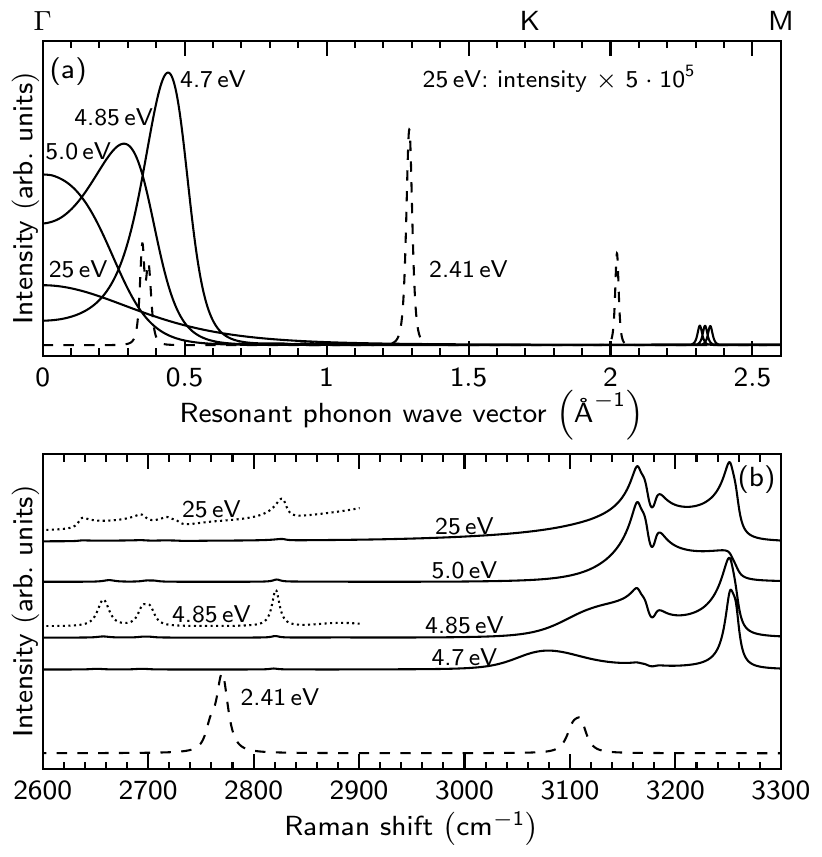}
\caption{(a) Resonant phonon wave vectors in graphite of an one-dimensional DR Raman process according to Eq.~\eqref{eq:DR} for different excitation energies given next to the curves. The intensity for $25\,\text{eV}$ excitation energy is multiplied by $5\cdot 10^{5}$. Resonant 'outer' phonon wave vectors with $q\sim 2.3/$\AA\hspace{1mm} correspond to the low-intensity Raman bands shown by dotted lines in (b). (b) Calculated Raman spectra according to Eq. (1). Dotted lines show parts of the spectra multiplied by $10^{2}$. All spectra are normalized and vertically offset for clarity. \label{fig4}}
\end{figure}

\begin{figure*}
\includegraphics[width=2\columnwidth]{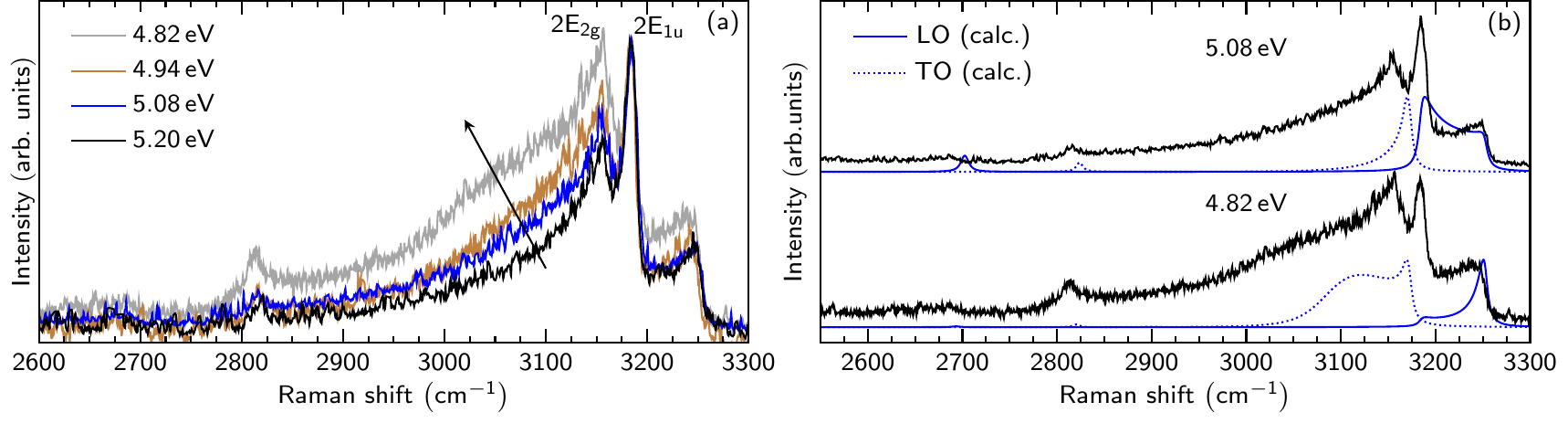}
\caption{(a) Raman spectra of graphite for different excitation energies. The evolution of a broad and dispersive peak at $3050\,\text{cm}^{-1}$ can be clearly observed for decreasing laser energies. For better visualization, all spectra are normalized to the overtone of the non-resonant, infrared-active $E_{1u}$ mode. (b) Comparison of experimental (black) and calculated (blue) Raman spectra of graphite for two excitation energies. Dashed (solid) lines correspond to the TO (LO) phonon branches. \label{fig5}}
\end{figure*}

We will now turn our discussion to the DR Raman scattering process in the UV range. Figure~\ref{fig3}\,(a) shows a contour plot of equi-energy lines from the electronic band structure of graphene with $E_{c}-E_{v}=\text{constant}$ ($E_{c}$: conduction band; $E_{v}$: valence band). We highlight the contour where $E_{c}-E_{v}= 5.08\,\text{eV}$, \textit{i.e.}, the equi-energy line where the absorption for an excitation wavelength of $244\,\text{nm}$ occurs. In contrast to excitation in the visible range, UV-excitation contours form circles around the $\Gamma$ point rather than closed triangles around the $K$ point. Furthermore, using excitation energies above the $M$-point transition energy, we can selectively suppress optical transitions along the $K-M$ high-symmetry direction in the double-resonance process [compare Figure~\ref{fig3}\,(b)]. These transitions are so-called 'inner' processes, whereas excitations along the $K-\Gamma$ direction with phonons from $K-M$ are 'outer' processes. However, due to low intensity, the latter contribute only marginally to the Raman spectrum in the double-resonance process \cite{Venezuela2011, Herziger2014}. Therefore, the UV spectra shown in Fig.~\ref{fig2} are dominated by the (non-resonant) two-phonon pDOS signal. Tuning the excitation energy from the deep UV to $4.7\,\text{eV}$, allows optical transitions from the $K-M$direction and thus activates dominant inner processes.
 
In the following, we present and discuss calculated Raman spectra of graphite for excitation energies between $25\,\text{eV}-4.7\,\text{eV}$ in the spectral range from $2600\,\text{cm}^{-1}$ to $3300\,\text{cm}^{-1}$ [see Fig.~\ref{fig4}\,(b)]. DR Raman spectra were calculated using the equation \cite{Thomsen2000}
\begin{align}
\mathcal{I} \propto	\sum_{\alpha=1}^{6}\bigg\vert\sum_{a,b,c}&\frac{\mathcal{M}}{(E_{L}-E_{ai}-i\gamma)(E_{L}-E_{bi}-\hbar\omega_{\alpha}-i\gamma)} \notag\\
					\times&\frac{1}{(E_{L}-E_{ci}-2\hbar\omega_{\alpha}-i\gamma)}\bigg\vert^{2}, \label{eq:DR}
\end{align}
where we assumed the matrix elements $\mathcal{M}$ to be constant. $E_{xi}$ denote the energy differences between the intermediate electronic states $a,b,c$ and the initial state $i$. We considered the six highest phonon branches (LO, TO, LA), which are indexed with $\alpha$. The integration is performed along the $\Gamma-K-M$ high-symmetry line, including the scattering of both electrons and holes\cite{Venezuela2011}. The broadening factor $\gamma$ was set to $120\,\text{meV}$ for all excitation energies \cite{Venezuela2011}. The electronic bands are GW corrected, in order to fit the experimentally observed $M$ point energy. \cite{Mak2011}

The calculated spectrum in Fig.~\ref{fig4}\,(b) for an artificially high excitation energy of $E_{L}=25\,\text{eV}$ resembles the pDOS of graphite, since only off-resonant contributions dominate the Raman spectrum. By decreasing the excitation energy to values of $5\,\text{eV}$ or less, inner DR processes with phonons exhibiting ultra-short wave vectors are activated and thus start to dominate the spectra [Fig.~\ref{fig4}\,(a)]. They mediate DR scattering processes between two electronic states close to the $M$ point. Due to the flat slope of the electronic bands around the $M$ point, the double-resonance process can be fulfilled by a broad range of phonon wave vectors. This directly results in broad peaks in the Raman spectra around $3100\,\text{cm}^{-1}$, as can be seen in Fig.~\ref{fig4}\,(b). A further decrease of the excitation energy results in a smaller peak width and a downshift of the DR Raman peak, corresponding to the slope of the electronic bands and TO/LO phonon branches. As can be seen, the relative intensity of the LO-overbending Raman peak increases with lower excitation energies, since the resonant phonon wave vector approaches the maximum of the LO phonon branch. The maximum exhibits a wave vector of 0.46/\AA{} and can be resonantly accessed by excitation energies marginally lower than $4.7\,\text{eV}$ (Fig.~\ref{fig4}). 

In Figure~\ref{fig4}\,(b), we also highlight the spectral range that is associated with 'outer' DR processes and exhibits several low-intensity Raman peaks. These peaks are due to 'outer' scattering processes with LA, LO and TO phonons and exhibit resonant phonon wave vectors with $q\approx 2.3/\text{\AA}$ [Fig.~\ref{fig4}\,(a)]. Thus, we can observe phonons with large wave vectors close to the $M$ point using excitations in the deep-UV. For $E_{L}=25\,\text{eV}$, the lineshape of the low-intensity peaks directly corresponds to the pDOS, whereas they transform into Lorentzian lineshapes for lower excitation energies, since they correspond to outer DR processes. In general, the contributions from the pDOS-related Raman bands drastically decrease at lower excitation energies, as the enhancement of the DR processes outperforms the pDOS-related peaks by orders of magnitude.

The experimental Raman spectra of graphite for four different excitation energies are presented in Fig.~\ref{fig5}. As predicted by the calculations (Fig.~\ref{fig4}), by lowering the excitation energy, we clearly observe the evolution of a broad and dispersive peak around $3050\,\text{cm}^{-1}$. This peak is the beginning of the \textit{2D} mode in graphite, as it is due to a double-resonance process at the $M$ point with two TO phonons exhibiting ultra-short wave vectors. As predicted theoretically, we further observe an intensity increase of the LO-overbending Raman peak for lower excitation energies. In Fig.~\ref{fig5}\,(b), we explicitly decompose the calculated spectra into contributions from the TO and the LO phonon branch in the double-resonance process for two different excitation energies. It shows the radical change in both the experimental and calculated Raman spectra mainly caused by the TO phonon branches (blue, dashed lines), when inner processes are activated. This fact emphasizes the importance of inner contributions in DR Raman processes in graphite and graphene.\newline 

In conclusion, we performed an analysis of deep-UV Raman spectra for graphite, graphene and CNT. At an excitation energy of $5.08\,\text{eV}$, we observe the phonon density of states in the second-order spectra of all samples. Decreasing the excitation energy towards the transition energy of the $M$ point, we can activate inner DR processes that are mediated by ultra-short phonon wave vectors. We observe the evolution of a broad and dispersive peak for excitation energies below 4.9\,eV, which is the rise of the \textit{2D} mode in graphite. Our calculated Raman spectra show good agreement with the experimental data and support our interpretation of the Raman process in the UV spectral range. Finally, our experimental findings can be helpful to fine-adjust calculations of the phonon dispersions of graphene and graphite.

\begin{acknowledgments}
This work was supported by the Deutsche Forschungsgemeinschaft (DFG) within SPP 1459 "Graphene" (MA 4079/7-2) and FOR1282 (MA 4079/6-2) and by the European Research Council (ERC) Grant No. 259286.
\end{acknowledgments}

\end{document}